\documentclass[journal]{IEEEtran}
\usepackage{amsmath,amsfonts}
\usepackage{algorithmic}
\usepackage{algorithm}
\usepackage{array}
\usepackage[caption=false,font=normalsize,labelfont=sf,textfont=sf]{subfig}
\usepackage{textcomp}
\usepackage{stfloats}
\usepackage{url}
\usepackage{verbatim}
\usepackage{graphicx}
\usepackage[normalem]{ulem}
\usepackage{cite}
\usepackage{xcolor}
\usepackage{gensymb}
\usepackage{multirow}
\usepackage[nolist,nohyperlinks]{acronym}
\begin{acronym}
    \acro{BS}{base station}
    \acro{MU}{mobile user}
    \acro{ISAC}{integrated sensing and communication}
    \acro{BER}{bit error rate}
    \acro{EVM}{error vector magnitude}
    \acro{RMSE}{root mean square error}
    \acro{mmWave}{millimeter wave}
    \acro{RSP}{radar signal processing}
    \acro{MFE}{millimeter wave front end}
    \acro{RF}{radio-frequency}
    \acro{IF}{intermediate frequency}
    \acro{IQ}{inphase-quadrature}
    \acro{LO}{local oscillator}
    \acro{CFO}{carrier frequency offset}
    \acro{ULA}{uniform linear array}
    \acro{SNR}{signal-to-noise ratio}
    \acro{Gbps}{gigabits per second}
    \acro{OFDM}{orthogonal frequency division multiplexing}
    \acro{FPGA}{field programmable gate array}
    \acro{RFSoC}{radio frequency system on chip}
    \acro{DFE}{digital front end}
    \acro{DAC}{digital-to-analog convertor}
    \acro{ADC}{analog-to-digital convertor}
    
\end{acronym}
\usepackage{hyperref}
\usepackage{dirtytalk}
\usepackage{pifont}
\usepackage{url}

\usepackage{lipsum}

\begin{document}

\title{Millimeter Wave Frontend for Integrated Sensing and Communication System Transceiver on Edge}

\author{Jai Mangal, Kshitiz Joshi*, Krishna Neel Reddy*, Soumya Jain, Shobha Sundar Ram, \textit{Senior Member, IEEE} and Sumit J. Darak,  \textit{Senior Member, IEEE} 
 \thanks{*Kshitiz Joshi and Krishna Neel Reddy are joint second co-authors.}
\thanks{This work is supported by the funding received from Chip to Startup (C2S), MeiTy, GoI and Qualcomm Innovation Fellowship (QIF) 2024, India. All authors are with ECE Dept., IIIT-Delhi, India. (e-mail: \{jaim, kshitiz19169, krishna19182, soumya20325, shobha, sumit\}@iiitd.ac.in}			
}

\maketitle
\begin{abstract}
IEEE 802.11ad standard uses analog beamforming for high-speed directional communication with \ac{MU} in the \ac{mmWave} spectrum. However, the lengthy beam alignment procedures involving large data packets between the \ac{BS} and the \ac{MU} introduce considerable overhead, deteriorating the overall throughput. Prior works have proposed 802.11ad-based \ac{ISAC} \ac{BS} transceivers to eliminate time-consuming beam alignment. Instead, the radar and communication functionalities use the same waveform, spectrum, and \ac{MFE} with a common spatial field of view. The radar detects and localizes the \ac{MU}, enabling the subsequent directional communication with the \ac{MU}. This work proposes an end-to-end IEEE 802.11ad-based \ac{ISAC} \ac{BS} transceiver prototype, wherein the digital baseband hardware frontend on edge is integrated with a Simulink-based \ac{MFE}.  The proposed prototype facilitates a systematic link budget and detailed performance analysis for different wireless channels, target motions, signal-to-noise ratios, hardware configurations, and impairments. We also investigate how these impairments affect radar performance and, in turn, the communication metrics since the performances of both systems are uniquely interrelated in an ISAC system. Our results show that even with hardware impairments, the 802.11ad-based ISAC offers 34\% higher throughput than the standard with an ideal \ac{MFE}.
\end{abstract}

\begin{IEEEkeywords}
millimeter wave front end, integrated sensing and communication, millimeter wave frequency, RF impairments, wideband system
\end{IEEEkeywords}

\section{Introduction}
\label{sec:Intro}
Vehicle-to-everything (V2X) communication supports the accurate sharing of real-time traffic and road condition information among road users and between vehicles and road infrastructure, enabling faster and more efficient transportation. 
Traditionally, 3GPP-based solutions such as Long Term Evolution-V2X (Release 14/15) and New Radio-V2X (Release 16/17) have been widely used for vehicular communications because they support long-range, cellular-based connectivity. One of the critical limitations of 3GPP-based V2X is the limited spectrum availability. LTE-V2X is typically constrained to 10-20 MHz bandwidth per channel, while NR-V2X (5G) extends this to a maximum of 400 MHz in the sub-6 GHz spectrum \cite{chen2017v2x}.
Therefore, \ac{mmWave} spectrum is being explored for high bandwidth ultra-reliable vehicular networks \cite{choi2016millimeter}. Specifically, the IEEE 802.11ad/ay standards that offer above 2 GHz bandwidth have been proposed to support single carrier \ac{mmWave} transmission at 4.6 \ac{Gbps} and \ac{OFDM} at 6.8 \ac{Gbps} \cite{ieee80211-2016}. The advantages of IEEE 802.11ad over 3GPP-based V2X solutions are summarized in Table \ref{tab:v2x_comparison}. However, unlike the sub-6 GHz carrier frequencies, the \ac{mmWave} frequency is characterized by higher atmospheric attenuation requiring narrow directional beams for transmission \cite{liu2017millimeter}. Hence, in vehicular environments, both the \ac{BS} and \ac{MU} must either, independently or jointly, rapidly determine the optimal beams for communication. 

\begin{table*}[h]
    \centering
    \renewcommand{\arraystretch}{1.2}
     \caption{Comparison of IEEE 802.11ad and 3GPP-based V2X Standards}
    \begin{tabular}{|p{3cm}|p{4.5cm}|p{4.5cm}|p{4.5cm}|}
        \hline
        \textbf{Feature} & \textbf{IEEE 802.11ad (WiGig)} & \textbf{LTE-V2X (Rel. 14/15)} & \textbf{NR-V2X (Rel. 16/17)} \\
        \hline
        Bandwidth (BW) & 1.76 or 2.64 GHz & 10 to 20 MHz, 100 MHz (with carrier aggregation) & 400 MHz, 700 MHz (with carrier aggregation) \\
        \hline
        Maximum Data Rate & Up to 7 Gbps & ~100 Mbps & Up to 1 Gbps \\
        \hline
        Latency & 1 ms & ~10 ms & ~1 ms \\
        \hline
        Communication Type & Direct V2V, V2I (ad-hoc) & Network-assisted V2V, V2I, V2N & Network-assisted V2V, V2I, V2N \\
        \hline
        Range & Short-range (~200m) & Long-range (~1 km) & Long-range (~1 km) \\
        \hline
        Range resolution & 0.085m  & 7.5m & 0.37m\\
        \hline
        Beamforming & Yes (Highly directional) & No (Omnidirectional) & Yes (for mmWave bands) \\
        \hline
    \end{tabular}
    \label{tab:v2x_comparison}
\end{table*}

Several strategies have been explored to facilitate the beam alignment. Standard IEEE 802.11ad/ay protocols enable the scanning across multiple narrow candidate beams at both \ac{BS} and \ac{MU} \cite{ieee80211-2016}. However, this leads to lengthy overheads and poor latency, significantly deteriorating communication throughput. Alternatively, researchers have studied the possibility of using auxiliary sensors, such as standalone radars, for localizing the \ac{MU} to avoid the requirement of continuous beam scanning and realignment. However, this strategy faces the challenges of managing interference and synchronization between the communication and auxiliary sensors \cite{aviles2016position}. Another approach that researchers have investigated involves integrating the communication system with radar functionality. This integration allows both systems to share hardware and spectral resources and avoid synchronization and interference issues. Consequently, this has spurred the research of new waveforms \cite{knill2018high}, modulation schemes \cite{oliveira2022joint}, signal processing algorithms \cite{liu2022survey}, transceiver components \cite{alam2022integrated} and architectures\cite{liu2022integrated,sneh2023radar,sneh2024beam}. 

The earliest work on \ac{mmWave} IEEE 802.11ad based \ac{ISAC} was by \cite{kumari2018ieee}, where the Golay sequence in the communication protocol's preamble was noted for its suitability for radar ranging due to its perfect autocorrelation properties and high bandwidth. The authors in \cite{duggal2019micro,duggal2020doppler} then proposed modifying the Golay sequences to enhance its Doppler resilience to facilitate the detection of \ac{MU}.  In our preliminary work in \cite{sneh2024ieee}, we presented the architecture for an 802.11ad-based \ac{ISAC} along with the software prototype of the digital front end of \ac{BS}. Further, we presented the \ac{ISAC} \ac{RSP} accelerator realized on a multi-processor system on chip (MPSoC) in \cite{tewari2025reconfigurable}. Other \ac{ISAC} setups are developed at 28 GHz in \cite{zhang2021design}, with maximum detectable radar range around 10m and above 2 Gbps data rates. Recent works that have presented software/hardware prototypes for \ac{ISAC} system at \ac{RF} and \ac{mmWave} frequencies are listed in Table \ref{tab:LR}. 
Several of these works focus only on the radar or communication metrics separately but not on how these metrics are interrelated. Specifically, since spectrum and hardware resources are shared between the two functionalities, the performances of sensing and communication are not independent of each other. Instead, based on the nature of hardware configuration and design objectives, the sensing performance may augment the communication performance or vice versa. In this work, we focus on this particular unique aspect of ISAC. 

Finally, \ac{ISAC} transceivers will be an integral part of the distributed units (DU) and radio units (RU) of the wireless networks. The DU and RU are remotely deployed, and hence, their efficient implementation on resource-constrained edge platforms such as processor, \ac{FPGA} or application specific integration circuits (ASIC) is desired. Existing prototypes do not present a hardware implementation of the digital frontend (DFE) in either floating or fixed point arithmetic. The functionally accurate fixed-point architecture is critical for lowering area, delay and power footprint. While some works consider analog or digital beamforming architecture in the \ac{MFE}, none consider the effects of realistic RF impairments of these hardware components on the \ac{ISAC} performance.

\begin{table*}[!b]
\caption{Summary of literature survey comparing the contributions of the proposed work with the prior art for \ac{ISAC}
\label{tab:LR}}
\centering
\begin{tabular}
{|c|c|c|c|c|c|c|c|c|c|c|} \hline
\textbf{Ref. No.} & \multicolumn{3}{|c|}{\textbf{Performance Metrics}} & \multicolumn{3}{|c|}{\textbf{Digital Front-end}} & \multicolumn{4}{|c|}{\textbf{Analog Front-end}} \\ \hline
 & \textbf{Radar} & \textbf{Comm.} & \textbf{\ac{ISAC}} & \textbf{Software} & \multicolumn{2}{|c|}{\textbf{Hardware}} & \textbf{Prototype} & \multicolumn{2}{|c|}{\textbf{Beamforming}} & \textbf{Impairments} \\ \hline
& & & & & \textbf{Float} & \textbf{Fixed} & & \textbf{Radar} & \textbf{Comm.} & \\ \hline
\textbf{\cite{sneh2024ieee}} & \ding{51} & \ding{51} & \ding{51} & \ding{51} & \ding{55} & \ding{55} & \ding{55} & Digital & Analog & \ding{55} \\ \hline
\textbf{\cite{tewari2025reconfigurable}} & \ding{51} & \ding{55} & \ding{55} & \ding{51} & \ding{51} & \ding{51} & \ding{55} & Digital & None & \ding{55} \\ \hline
\textbf{\cite{bai2022photonic}} & \ding{51} & \ding{51} & \ding{55} & \ding{55} & \ding{55} & \ding{55} & \ding{51} & None & None & \ding{55} \\ \hline
\textbf{\cite{zhang2021performance}} & \ding{51} & \ding{51} & \ding{55} & \ding{51} & \ding{55} & \ding{55} & \ding{51} & None & None & \ding{55} \\ \hline
\textbf{\cite{zhang2021design}} & \ding{51} & \ding{51} & \ding{55} & \ding{51} & \ding{55} & \ding{55} & \ding{51} & None & None & \ding{55} \\ \hline
\textbf{\cite{kumari2018ieee}} & \ding{51} & \ding{55} & \ding{55} & \ding{51} & \ding{55} & \ding{55} & \ding{51} & Digital & None & \ding{55} \\ \hline
\textbf{\cite{ozkaptan2024mmwave}} & \ding{51} & \ding{51} & \ding{51} & \ding{51} & \ding{55} & \ding{55} & \ding{51} & Digital & None & \ding{55} \\ \hline
\textbf{\cite{ozkaptan2023software}} & \ding{51} & \ding{51} & \ding{51} & \ding{51} & \ding{55} & \ding{55} & \ding{51} & Digital & None & \ding{55} \\ \hline
\textbf{Proposed Work} & \ding{51} & \ding{51} & \ding{51} & \ding{51} & \ding{51} & \ding{51} & \ding{51} & Digital & Analog & \ding{51} \\ \hline
\end{tabular}
\end{table*}

The main contributions of this paper are as follows:\\
\textbf{(1) Complete Implementation of \ac{DFE}:} This paper presents a complete implementation of an IEEE 802.11ad-based ISAC with RSP on FPGA. The design includes custom Intellectual Properties (IPs) tailored for both radar and communication system applications and optimized for hardware efficiency. This optimization is achieved through fine-tuning the word length for fixed-point representation and employing serial-parallel processing techniques to balance speed and resource utilization effectively. The precision and computational efficiency of the fixed-point architecture are further enhanced by utilizing dynamic word-length scaling based on the computational demands at each stage of processing. Our design demonstrates a 166\% improvement in throughput compared to the baseline IEEE 802.11ad standard. 
\noindent\textbf{(2) Reconfigurable Hardware Architecture:} In this work, we introduce a novel reconfigurable hardware architecture that provides dynamic, on-the-fly control over the angular precision in RSP. This reconfigurability enables the system to adapt to varying operational conditions and optimize resource allocation in real time. We investigate the trade-offs between angular estimation precision and its impact on communication performance.
Our analysis shows that reduced angular precision can lead to faster RSP, which, in turn, results in more available time for communication effectively boosting throughput. 
\noindent\textbf{(3) Link budget analysis:} We present a detailed link budget analysis for the proposed radar system, determining the maximum detectable radar range to be 41.22 meters at a minimum \ac{SNR} of 2.28 dB. 
These metrics align with existing ultra-short-range automotive radars. 
\noindent\textbf{(4) Impact of Hardware Impairments on System Performance:} In this work, we systematically analyze the impact of hardware impairments on the functional performance of both the sensing and communication subsystems. Specifically, we evaluate how RF impairments, such as \ac{IQ} gain and phase imbalance, \ac{LO} leakage, and \ac{CFO}, affect the system's performance during target motion along both radial and tangential trajectories. For performance benchmarking, the throughput of the proposed \ac{ISAC} system with an ideal \ac{MFE} is 0.77 Gbps for the radial trajectory and 0.66 Gbps for the tangential trajectory. However, when accounting for RF impairments introduced by a non-ideal \ac{MFE}, we observe a degradation in throughput of 8.57\% and 6.61\% for the radial and tangential trajectories, respectively. 
Despite this degradation, the ISAC system, based on the IEEE 802.11ad standard, still delivers 34.84\% higher throughput for radial trajectories and 50.12\% higher throughput for tangential trajectories compared to the IEEE 802.11ad standard using an ideal \ac{MFE}. 
\noindent \textbf{(5) \ac{ISAC} Metrics:} In this work, we introduce a joint metric framework for both radar and communication systems to quantify and demonstrate the impact of radar on communication performance within the \ac{ISAC} system. 
Specifically, we explore the impact of radar-induced angular errors on communication system performance. 
These metrics highlight the crucial role of radar in maintaining communication link quality and provide a comprehensive understanding of the interdependence between radar sensing and communication in an \ac{ISAC} system. 

We organize the paper as follows. In Sections \ref{Sec:SM}, we present a high-level overview of the 802.11ad based \ac{ISAC} system model followed by the signal model. The proposed architecture of the \ac{DFE} on \ac{FPGA} with an ideal \ac{MFE} is presented in Section \ref{Sec:PA_DFE} and the corresponding performance and complexity analysis are presented in Section \ref{sec:Results_DFE}. This is followed by the proposed architecture of the \ac{MFE} with realistic RF impairments in Section \ref{sec:PA_MMFE}. The performance analysis of the effects of these impairments is presented in Section \ref{sec:Results_MMFE} followed by the conclusion in Section \ref{sec:Conclusion}.

\emph{Notation:} We represent scalar variables, vectors, and matrices with lowercase ($x$), bold lowercase ($\mathbf{x}$), and bold uppercase ($\mathbf{X}$) letters, respectively. We denote the transpose operation with $(\cdot)^T$ and the conjugate transpose operation with $(\cdot)^{\ast}$. Time domain and frequency domain versions of a vector are represented by $\mathbf{x}$ and $\tilde{\mathbf{x}}$, respectively.

\section{IEEE 802.11ad based ISAC System Model}
\label{Sec:SM}
The IEEE 802.11ad standard's packet consists of the preamble (P) followed by header (H), data (D), and $N$ optional beam refinement fields (BRF). Here, the P is modulated with a single carrier (with 1.76 GHz sampling frequency) and is transmitted at the beginning of each packet for synchronization and channel estimation to enable reliable reception of H and D. The H and D are \ac{OFDM} to achieve high data rate (with 2.64 GHz sampling frequency). High atmospheric attenuation in the \ac{mmWave} spectrum demands directional communication links, which means \ac{BS} and \ac{MU} must identify the appropriate beam direction in the initial beam alignment phase before data communication.  P, H, and D are transmitted via an omnidirectional beam, while BRFs are transmitted in $N$ pre-determined distinct directions via analog beamforming. 
\begin{figure}[!b]
    \centering
    \includegraphics[scale=0.67]{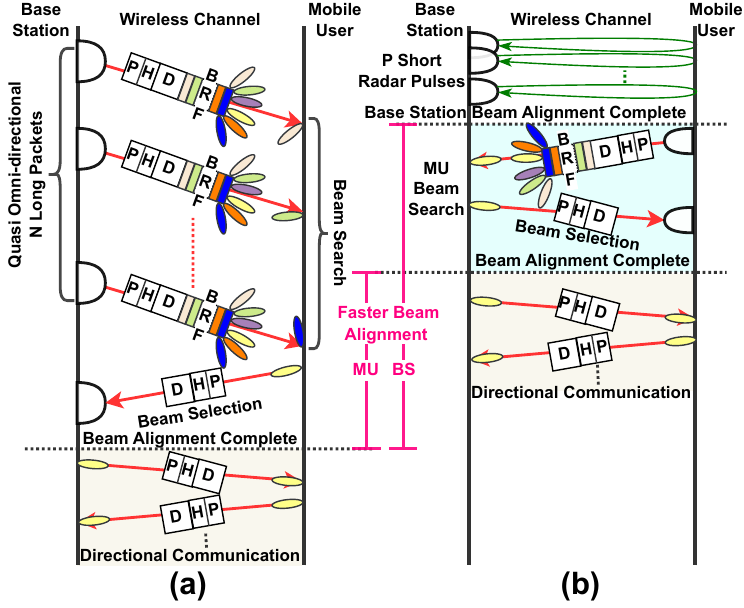}
    \vspace{-3mm}
    \caption{ Beam alignment procedures in (a) IEEE 802.11ad standard, and (b) IEEE 802.11ad based \ac{ISAC}.}
    \label{fig:BA}
    \vspace{-3mm}
\end{figure}
For instance, consider a \ac{BS} with $N$ beams and an \ac{MU} with $L (< N)$ beams. As shown in Fig.~\ref{fig:BA}(a), the initial beam training and realignment (BTR) phase in existing IEEE 802.11ad involves the transmission of $L$ long downlink (DL) packets from the \ac{BS}. Each of the $L$ packets is received through a distinct pre-determined beam at the \ac{MU} through analog beamforming. The \ac{MU} processes all $L$ received packets and selects the optimum beam pair from the \ac{BS} and \ac{MU} based on the received signal strength. The \ac{MU} communicates the selected beam index to \ac{BS} via an uplink packet, completing the BTR phase. For multiple MUs, each \ac{MU} sends separate uplink packets to communicate the optimum beam index. \\
\indent Our proposed beam alignment process with the \ac{ISAC} framework was presented in detail in \cite{sneh2024ieee} and is briefly recapitulated here with the help of the illustration in Fig.\ref{fig:BA}(b). Here, we time division multiplex between the radar and communication functionalities. First, we generate $P$ short radar pulses extracted from the preamble of 802.11ad and transmitted through omnidirectional beams from a single dipole transmit antenna with pulse repetition interval, $T_{PRI}$. The radar receiver at the BS consists of a separate antenna array capable of digital beamforming. 
We process the scattered signals from the MU to estimate the best beam from the BS. 
The MU does not have sensing capability. Therefore, the best beam selection at the MU takes place according to the principles of the standard 802.11ad using $L$ BRF, each transmitted along a distinct beam. Then, the BS determines the optimal beam based on the highest signal strength and communicates the information to the MU in the subsequent packet. 
Note that this does not introduce significant overhead since the number of candidate beams at \ac{MU} is relatively low (compared to \ac{BS}). The overall beam alignment duration for the modified ISAC-based 802.11ad is lower than the standard 802.11ad due to (1) the short radar pulses, which do not need decoding compared to the long communication packets used in the standard, and (2) the direct estimation of the angular position of the \ac{MU} through digital beamforming instead of the time-consuming beam by beam search in the standard using analog beamforming.

Once this stage is completed, communications follow through directional beams but with reduced packet size since the BRF is omitted. The \ac{MFE} with the antenna array (used earlier in the radar receiver) is used in a half-duplex mode in the communication scenario such that both the transmitter and receiver of the BS communication system time share the antenna array.
Further, we introduce a radar-communication decision unit to distinguish the radar-scattered signal from the communication uplink. With multiple MUs, each \ac{MU} sends separate uplink packets simultaneously, which \ac{BS} receives via digital beamforming. In the end, \ac{BS} communicates the selected beam index to each \ac{MU} via separate DL in a time-division multiplexed manner. 
\subsection{802.11ad based ISAC Signal Model}
\label{Sec:sig_mod}
\subsubsection{Radar signal and signal processing} 
Each 802.11ad communication packet, comprising P, H, and D, is denoted by $\mathbf{y}$. The transmitted radar signal is extracted from a $K$ length Golay sequence within the 802.11ad packet preamble and denoted by $\mathbf{x}$ with $T_s$ sampling time. The sequence is converted from digital to analog in the \ac{DAC}, upconverted to carrier frequency $f_c$ (and wavelength, $\lambda$), and transmitted from the \ac{BS} through an omnidirectional antenna. We consider an $m^{th}$ \ac{MU} located at polar coordinates ($r_m,\phi_m$) and moving with radial velocity $v_m$. The radar echo from $M$ \ac{MU} received at the \ac{BS}, consisting of a $Q$-element \ac{ULA} with $d$ inter-element spacing after downconversion and digitization, is 
\par\noindent\small
\begin{equation}
\label{eq:received_sig}
\begin{split}
  \mathbf{X}_{{rx}}[k,q] = \sum_{m=1}^Ma_{m}\mathbf{x}[k-k_{m}]e^{-j 2\pi f_{m}kT_s}e^{j \frac{2\pi}{\lambda}dq\sin\phi_m} + \eta.
  \end{split}
\end{equation} 
\normalsize 
Here, the amplitude, $a_{m}$, incorporates the scattering coefficient of the \ac{MU} and the two-way line-of-sight (LOS) propagation between the \ac{BS} and \ac{MU} and $\eta$ is the complex additive white Gaussian noise (AWGN). The received radar echo consists of range-induced delay, $k_{m} = \frac{2r_{m}}{cT_s}$ where $c$ is the speed of light, and Doppler-induced frequency shift, $f_{m} = \frac{2v_{m}}{\lambda}$. Note that $k_{m}$ and $f_{m}$ are unknown to the receiver and are estimated through \ac{RSP} at the \ac{BS} receiver. The received signal shown in \eqref{eq:received_sig} models only the LOS propagation. To account for multipath and clutter, we modify it as

\begin{align}
\label{eq:RxSig_ric}
\begin{split}
\mathbf{X}_{rx}[k,q]=  \sum_{m=1}^M\left[a_{m}\sqrt{\frac{\mathcal{J}}{\mathcal{J}+1}} + \rho\sqrt{\frac{1}{\mathcal{J}+1}}\right] \\ \mathbf{x}[k-k_{m}]e^{-j 2\pi f_{m}kT_s}e^{j \frac{2\pi}{\lambda}dq\sin\phi_m} + \beta.
\end{split}
\end{align}
\normalsize

Here, we model the contributions of multipath versus LOS returns with the Rician factor, $\mathcal{J}$ and the strength of the multipath signal or clutter, $\rho$, is an independent and identically distributed complex Gaussian random variable with zero mean and unit variance. The first term in \eqref{eq:RxSig_ric} indicates the dominant LOS component, while the second term shows the additional multipath components \cite{yan2020channel}. The higher the value of $\mathcal{J}$, the lower the multipath contribution. Therefore, typically highway road conditions are modeled with high $\mathcal{J}$ while urban road junctions are modeled with low $\mathcal{J}$. The channel model equations are implemented in MATLAB and integrated into Simulink.

The input to the \ac{RSP} block is the digitized received signal: $\mathbf{X}_{rx}[k,q]$, which comprises $K$ fast time samples across $Q$ antenna elements. Each $q^{th}$ column of the fast-time samples is a time-delayed version of the transmitted Golay sequences, $\mathbf{x}$. Therefore, when each $q^{th}$ column of $\mathbf{X}_{rx}$ is correlated in time-domain/match filtered with $\mathbf{x}$, we obtain a peak corresponding to the range-induced time delay of the \ac{MU} ($k_m$) and zero sidelobes elsewhere due to perfect autocorrelation properties. Since correlation operation as shown in \cite{liu2014alldigital} is computationally complex, we use the matched filtering through the multiplication operation in the frequency domain by first converting the time-domain received signal for each $q^{th}$ antenna to the frequency domain through fast Fourier transform: $\hat{\mathbf{X}}_{{rx}}[k_f,q] = \text{1D-FFT}\{\mathbf{X}_{{rx}}[k,q], q = 1 \cdots Q\}$. To jointly estimate the range and azimuth, we perform digital beamforming across the $Q$ element array in conjunction with matched filtering as 
\par\noindent\small
\begin{align}
\label{eq:MF_FFT}
\hat{\mathbf{\chi}}[\phi_i] = \text{diag}\left[\hat{\mathbf{X}}_{{rx}}\mathbf{w}_{\phi_i}\hat{\mathbf{x}}^\ast \right].
\end{align} \normalsize
Here,  $\mathbf{w}_{\phi_i}  = [1\; e^{-j k_c d \sin\phi_i} \cdots e^{-j (Q-1)k_c d\sin \phi_i}]^T$ is the weight vector applied to the \ac{ULA} receiver data to scan the beam at every $\phi_i$, $i = 1 \cdots I$ 
and $\hat{\mathbf{x}}$ is the frequency domain version of the transmitted signal $\mathbf{x}$. 
The entire radar search space is spanned through increments in $\Delta \phi$. In \eqref{eq:MF_FFT}, $\text{diag}$ extracts the diagonal elements of the complex $K \times K$ matrix generated by the matrix multiplication operation. Finally, the output in the time domain is obtained using inverse FFT (IFFT) as: ${\mathbf{\chi}}[r_k,\phi_i] = \text{IFFT}\{\hat{\mathbf{\chi}}[k_f,q]\}$ where, $r_k$ denotes the range samples which is directly related to the $k^{th}$ fast time sample. The peak search in $\mathbf{\chi}$ gives the range and azimuth of the \ac{MU}.
\subsubsection{Communication signal and signal processing}
Once the \ac{MU} is localized, the directional communication phase starts, as shown in Fig.~\ref{fig:BA}. Here, the DL packet denoted by $\mathbf{y}$, which comprises $B$ samples, is assembled and then converted to analog through the Dirac delta function, $\delta(\cdot)$. These analog signals are then upconverted to $f_c$ and transmitted along $\phi_m$ direction through analog beamforming by the \ac{BS} through the application of the antenna weight vector, $\mathbf{w}_{\phi_m} \in \mathcal{C}^{N_{BS}\times 1}$. The communication signal transmitted at \ac{BS} is given as
\par\noindent\small
\begin{align}
\mathbf{Y}_{tx}(\tau) = \mathbf{w}_{\phi_m} \left(\sum_{b=0}^{B-1}\mathbf{y}[bT_s]\delta(\tau - bT_s)\right) e^{+j2\pi f_c \tau},
\end{align}
\normalsize
The $m^{th}$ \ac{MU} receiver with a $Q'$ element \ac{ULA} performs analog beamforming with antenna weight vector, $\mathbf{w}_{\phi'}\in \mathcal{C}^{Q'\times 1}$, followed by digitization and downconversion as
\par\noindent\small
\begin{align}
\begin{split}
\label{eq:1DProp}
\mathbf{Y}_{rx}[b] = \mathbf{w}^T_{\phi'}\mathbf{u}_{\phi'}\mathbf{u}_{\phi_m}^T\mathbf{Y}_{tx}[b-\frac{k_m}{2}] e^{-j2\pi \frac{f_m}{2} bT_s} + \zeta.
\end{split}
\end{align}
\normalsize
Here, the signal is time delayed by $k_m/2$ and frequency shifted by $f_m/2$ due to one-way propagation from \ac{BS} to \ac{MU}. $\mathbf{u}_{\phi_m}$ and $\mathbf{u}_{\phi'_m}$ are the complex steering vectors at the \ac{BS} and the \ac{MU}, respectively, and $\zeta$ is the complex AGWN at the \ac{MU} receiver. Similar to the radar signal model, we modify \eqref{eq:1DProp} by introducing the Rician factor $\mathcal{J}$ to model multipath in the received signal. Receiver PHY processes the received signal at the \ac{MU}, and the corresponding uplink is sent to the \ac{BS}. The uplink is distinguished from the radar received signal at the \ac{BS} receiver through cross-correlation with the transmitted radar signal $\mathbf{x}$. Cross-correlation between matched Golay sequences ensures a high peak-to-sidelobe ratio for radar-received signal compared to the uplink communication signal with a mismatched Golay sequence.
\section{Proposed Architecture of ISAC PHY and \ac{DFE} with Ideal \ac{MFE}}
\label{Sec:PA_DFE}
\begin{figure}[!ht]
    \centering
    \includegraphics[scale=0.65]{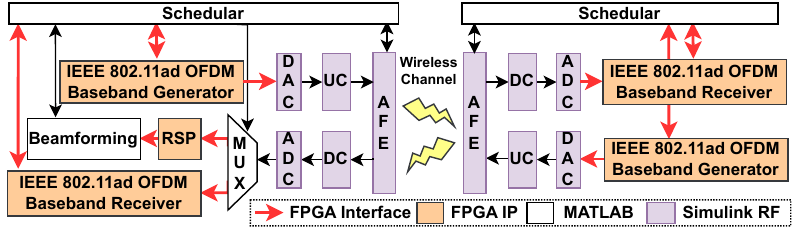}
    \caption{Block diagram of proposed IEEE 802.11ad based \ac{ISAC} system.}
    \label{fig:BD}
\end{figure}

The proposed 802.11ad-based \ac{ISAC} architecture involves the design of the hardware IPs for the transmitter and receiver PHY capable of transmitting and receiving the IEEE 802.11ad data frame. It also includes the hardware IP for \ac{RSP} to detect the \ac{MU} and estimate its range and azimuth to the \ac{BS}. For end-to-end \ac{ISAC} in Fig.~\ref{fig:BD}, the scheduler, which controls BTR and directional communication phases, data converters, \ac{MFE}, up-converter, down converter, wireless channels, and radar target modeling, are realized in the software using Matlab and interacts with the hardware IPs via \ac{FPGA}-based edge platform. As shown in Fig.~\ref{fig:BD}, impairments due to non-ideal \ac{ADC}, \ac{DAC}, and \ac{MFE} are part of the studies presented in Section~\ref{sec:PA_MMFE} onwards. 
The proposed framework is developed using various MATLAB and Simulink toolboxes to ensure a comprehensive ISAC system design. Specifically, the scheduler and baseband implementation of the ISAC transmitter, receiver, and \ac{RSP} leverage the Wireless, Radar, and Signal Processing toolboxes. The FPGA-based implementation of baseband algorithms, integrated via an FPGA-in-loop configuration, is achieved using the HDL Coder and Embedded Coder toolboxes. As discussed later in Section~\ref{sec:PA_MMFE}, the design of the mmWave front-end (MFE) utilizes the RF, Phased Array, Radar, and Antenna toolboxes. The proposed framework carefully integrates these toolboxes to enable an end-to-end ISAC implementation. To facilitate a better understanding and reproducibility of our design, we have provided a detailed tutorial explaining the framework. The tutorial and source codes are available \href{https://github.com/Dr-Jai-Mangal/MMW_ISAC.git}{here}.

\subsection{Transceiver PHY}
\label{Sec:TPHY}
\begin{figure}[!t]
    \centering
    \includegraphics[scale=0.57]{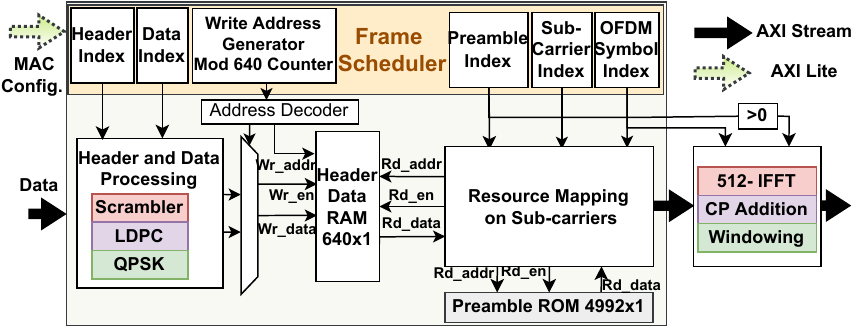}
    \caption{Architecture of 802.11ad based baseband transmitter PHY on \ac{FPGA}.}
    \label{fig:TX}
\end{figure}

The 802.11ad-based \ac{OFDM} transmitter PHY architecture for realization in \ac{FPGA} is shown in Fig.~\ref{fig:TX}. At the beginning of the data frame, P of size 4992 complex symbols is transmitted. It consists of an initial 32 samples followed by 15 repetitions of the next 192 samples and ending with 2080 samples. It is stored in read-only memory and sent directly to the data converters. The rest of the data frame comprises multiple \ac{OFDM} symbols where single H and multiple D are transmitted sequentially. To generate a single \ac{OFDM} symbol, input 512 bits are processed by a scrambler comprising of XOR operations followed by a low-density parity check (LDPC) encoder with 3/4 code rate involving a parity matrix of size 168$\times$504 to obtain 168 parity bits. These are appended to obtain an output of 672 bits, followed by data modulation to obtain 336 complex symbols. As per the 802.11ad standard, we have implemented tone pairing based QPSK modulation with a precoding matrix of size 2$\times$2. The 640 complex symbols corresponding to a single \ac{OFDM} symbol are buffered in memory. The scheduler then initiates the resource mapping of the complex symbols on appropriate sub-carriers along with pilots, followed by \ac{OFDM} modulation involving IFFT, cyclic prefix (CP) addition, and windowing. The frame scheduler generates appropriate control signals to transmit a data frame comprising P, H, and 10 data symbols. 

\begin{figure}[!t]
    \centering
    \includegraphics[scale=0.6]{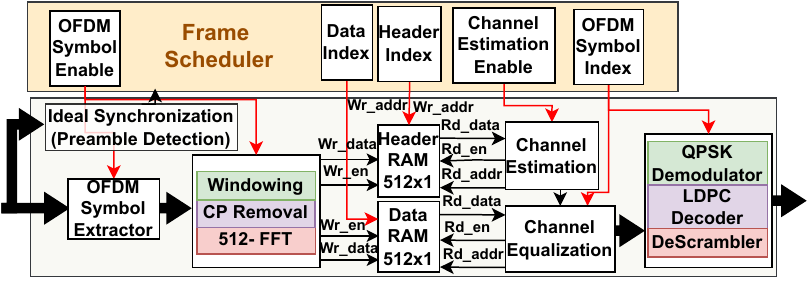}
    \caption{Architecture of 802.11ad based baseband receiver PHY on \ac{FPGA}.}
    \label{fig:RX}
\end{figure}

The corresponding 802.11ad-based \ac{OFDM} receiver architecture for realization in \ac{FPGA} is shown in Fig.~\ref{fig:RX}. We assume the receiver is synchronized with the transmitter to receive H and D symbols correctly. Then, \ac{OFDM} demodulation is done. The least square-based channel estimation is carried out on the received H, followed by minimum mean square error-based equalization on D. Like the transmitter, the QPSK demodulation, LDPC decoding, and descrambling is done followed by \ac{BER} analysis. 
\subsection{Radar Signal Processing }
\begin{figure}[!t]
    \centering
    \includegraphics[scale=0.7]{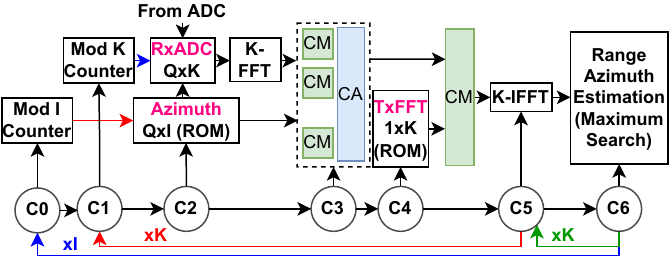}
    \caption{Architecture of range and azimuth estimation using \ac{RSP}.}
    \label{fig:RSP}
\end{figure}

The \ac{BS} receiver receives the scattered version of the transmitted radar pulses. Via digital beamforming and matched filtering, the digitized received signal is processed by the \ac{RSP} to obtain the range and azimuth of the \ac{MU} as discussed in the previous section. The \ac{RSP} is realized on the \ac{FPGA}, and the corresponding architecture is shown in Fig.~\ref{fig:RSP}. For $Q=32$ receive antennas, we capture $K=1024$ fast-time samples. Next, for each $\phi_i$ azimuth value, the matched filtering is done in the frequency domain using the sequentially transmitted reference waveform. This involves: 1) 1024-point fast Fourier transform (FFT) of fast-time samples, 2) Beamforming, which involves complex multiplication (CM) and complex addition (CA), 3) Match filtering via CM with reference waveform, and 4) Inverse FFT. In the end, we obtain a range-azimuth matrix of size $K \times I$, and the index corresponding to the maximum value gives the range and azimuth of the \ac{MU}. 
\section{Performance and Complexity Analysis of Proposed ISAC with Ideal MFE}
\label{sec:Results_DFE}
We present the experimental setup to evaluate the hardware prototype's performance and analyze its complexity. We consider the BS at the origin of a Cartesian space with the ground plane along $xy$ and height along $z$. The BS has a 32-element \ac{ULA} of half-wavelength antenna spacing. We consider a single isotropic point scatterer as \ac{MU} with a 4-element \ac{ULA}. Across multiple realizations, the location of the \ac{MU} is uniformly distributed in a 40m x 40m 2D Cartesian space while the target's velocity is uniformly varied uniformly from -30 m/s to 30 m/s. The \ac{MU}'s RCS is assumed to be 1$m^2$ similar to a pedestrian.
We consider both Friis' free space and  Rician channel models in our analysis. The Rician model includes multipath propagation along with LOS as per \eqref{eq:RxSig_ric} where the Rician factor ($\mathcal{J}$) can be adjusted to account for diverse environmental conditions including highways, semi-urban, and urban. In this work, $\mathcal{J}$ is chosen to be 5 dB. 

We compare the functional correctness in terms of RMSE and BER of the proposed fixed-point hardware architectures with a double-precision floating point (DPFP) based realization in Matlab and existing IEEE 802.11ad standard for different \ac{MU} trajectories and channels. 
Note that all of these results are generated with the assumption of an ideal MFE.
\subsubsection{Transceiver PHY}
\begin{figure}[!h]
    \centering
    \includegraphics[scale=0.35]{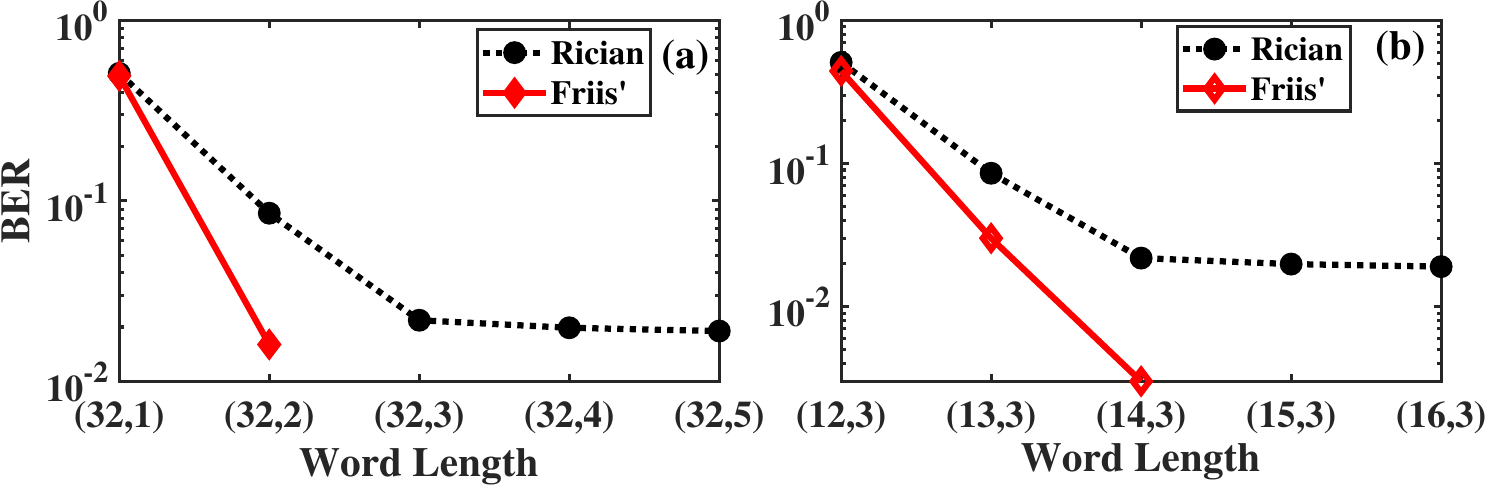}
    \caption{The \ac{BER} of the fixed-point transmitter PHY at SNR of 10 dB for different numbers of (a) Integer bits and (b) Fractional bits.}
    \label{fig:Tx_WL_JRC}
\end{figure}

For fixed-point architecture, we represent the wordlength (WL) as $(W,I)$ where $W$, $I$, and $(W-I)$ are the total number of bits, integer bits, and fractional bits, respectively. As shown in Fig.~\ref{fig:Tx_WL_JRC} (a), we first obtain a minimum value of $I$ for a sufficiently large $W$. It can be observed that three integer bits are sufficient for both channels as there is no improvement in BER with a further increase in $I$. Then, in Fig.~\ref{fig:Tx_WL_JRC} (b), we find the optimal value of $W$ by varying the number of fractional bits with $I=3$. It can be observed that $W=14$ is sufficient since there is no improvement in \ac{BER} after that. Thus, (14,3) is the preferred WL for the transmitter PHY. 

\begin{figure}[!t]
    \centering
    \includegraphics[scale=0.35]{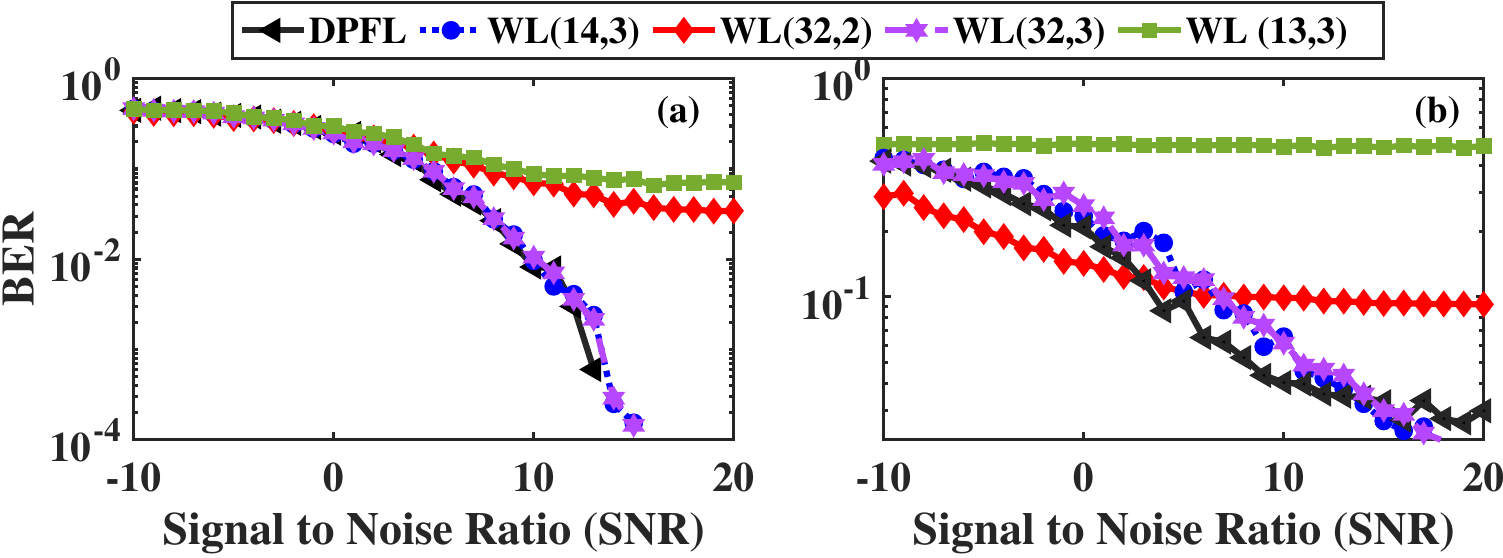}
    \caption{\ac{ISAC} transmitter \ac{BER} performance for different \ac{SNR}s and WLs in (a) Friis' channel, and (b) Rician channel.}
    \label{fig:Tx_channel_JRC}
\end{figure}

Figure~\ref{fig:Tx_channel_JRC} compares the \ac{BER} performance of the transmitter PHY for \ac{SNR} ranging from -10 dB to 20 dB and different WLs for the two wireless channels. Here, we assume the most accurate DPFP realization of the receiver PHY and \ac{RSP}. The \ac{BER} degrades with the \ac{SNR} decrease, and the Rician channel's \ac{BER} is poorer than Friis'. The proposed PHY with the WL of (14,3) achieves the same \ac{BER} performance as that of DPFP PHY realized in Matlab, validating the functional correctness of all the hardware IPs in transmitter PHY. Single bit reduction to WL of (13,3) results in significant degradation of \ac{BER}. Further, increasing WL beyond (14,3) does not improve the \ac{BER} performance as shown for the WL of (32,3). A larger WL of (32,2) with insufficient numbers of integer bits offers poor performance, and hence, careful selection of optimal WL is critical. 

Table~\ref{tab:tx_resource} shows that the resource and power consumption of WLs (13,3) and (14,3) are nearly identical, though the latter offers much lower \ac{BER}. Similarly, PHY with WL of (14,3) offers lower \ac{BER} and lower resource and power consumption than PHY with WL of (32,2). Among the two PHYs with nearly identical \ac{BER} performance, PHY with WL of (14,3) offers 33\%, 22\%, 31\%, and 59\% savings in slices, flip-flops, Block RAM, and DSP over PHY with WL of (32,3) along with 21\% lower power consumption.
\begin{table}[!ht]
\centering
\caption{Resource and Power Consumption of the Transmitter PHY on \ac{FPGA} for different WLs}
\label{tab:tx_resource}
\resizebox{\columnwidth}{!}{%
\begin{tabular}{|c|c|c|c|c|c|}
\hline
\textbf{WLs} & \textbf{\ac{FPGA} Slices} & \textbf{Flip-flops} & \textbf{Block RAMs} & \textbf{DSP} & \textbf{Power (W)} \\ \hline
\textbf{(13,3)} & 7096 & 27090 & 214 & 64 & 0.233 \\ \hline
\textbf{(14,3)} & 7184 & 27431 & 219.5 & 64 & 0.233 \\ \hline
\textbf{(32,2)} & 10874 & 35461 & 316.5 & 154 & 0.296 \\ \hline
\textbf{(32,3)} & 10923 & 35477 & 317.5 & 154 & 0.296 \\ \hline
\end{tabular}%
}
\end{table}

Similarly, we have analyzed the \ac{BER} of the receiver PHY for different WL and selected WL of (35,2). Due to space constraints and to avoid the repetition of discussion, we have skipped the corresponding results. 
The receiver demands higher WL than the transmitter due to challenging Rician channel conditions, which require accurate channel estimation. Furthermore, the receiver needs to perform soft data demodulation and LDPC demodulation, which demands higher WL. The receiver PHY with WL of (35,2) needs 42111, 24228, 202, and 288 flip-flops, slices, Block RAM, and DSP, respectively, with a power consumption of 0.44 W. 
\subsubsection{Radar Signal Processing}
We evaluate the functional performance of the \ac{RSP} by computing the \ac{RMSE} between the estimated range or azimuth, denoted as $\hat{a}$, and their ground truth counterpart values (denoted as $a$) over 100 experiments.
In each experiment, the range of the target is selected from a uniform distribution in the 2-D Cartesian space, with azimuth spanning from $-90^{\circ}$ to $90^{\circ}$; and the Doppler velocity spanning from $-30$ m/s to $+30$ m/s. 
Similar to transceiver PHY, we first obtain the desired WL, which offers performance similar to that of DPFL in Matlab. We observed that 
one integer bit is sufficient to achieve the same \ac{RMSE} as that of DPFL, and WL of (29,1) is sufficient for \ac{RSP}.

\begin{figure}[!t]
    \centering
    \includegraphics[scale=0.35]{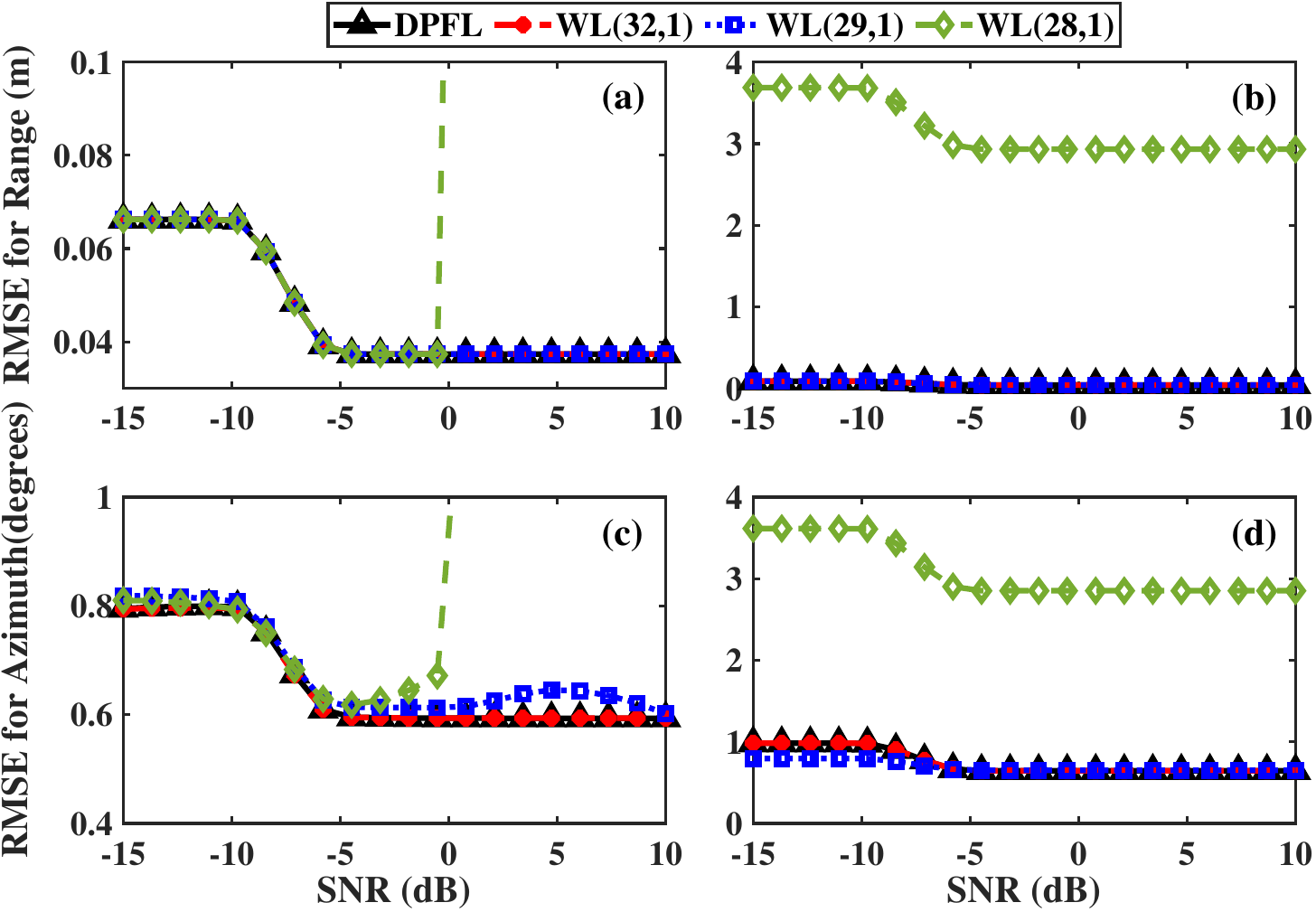}
    \caption{Performance analysis of \ac{RSP} on \ac{FPGA} for a wide range of \ac{SNR} and different WLs in terms of \ac{RMSE} for range in (a) Friis', and (b) Rician channels, and \ac{RMSE} for azimuth in (c) Friis', and (d) Rician channels.}
    \label{fig:RSP_JRC}
\end{figure}

Next, we validate the functional correctness of the \ac{RSP} on \ac{FPGA} for different WL over a wide range of \ac{SNR}s. As shown in Fig.~\ref{fig:RSP_JRC} (a)-(d), the \ac{RMSE} performance of fixed-point \ac{RSP} for WL of (29,1) is nearly identical to that of DPFL realization in Matlab for range and azimuth estimation. Single-bit reduction leads to significant degradation in \ac{RMSE}, especially for the Rician channel, and a higher WL of (32,1) does not offer any improvement in \ac{RMSE}. As shown in Table~\ref{tab:rsp_resource}, all three WLs consume nearly identical resources and power. 
\begin{table}[!ht]
\centering
\caption{Resource and Power Consumption of the \ac{RSP} for different WL}
\label{tab:rsp_resource}
\resizebox{\columnwidth}{!}{%
\begin{tabular}{|c|c|c|c|c|c|}
\hline
\textbf{WLs} & \textbf{\ac{FPGA} Slices} & \textbf{Flip-flops} & \textbf{Block RAMs} & \textbf{DSP} & \textbf{Power (W)} \\ \hline
\textbf{(28,1)} & 11792 & 26623 & 137 & 675 & 0.233 \\ \hline
\textbf{(29,1)} & 12209 & 27223 & 137 & 675 & 0.236 \\ \hline
\textbf{(32,1)} & 13065 & 28458 & 138 & 675 & 0.237 \\ \hline
\end{tabular}%
}
\end{table}
\subsubsection{End-to-End ISAC}
\label{Sec:ISACFPGA}
\begin{figure}[!ht]
    \centering
    \includegraphics[scale=0.73]{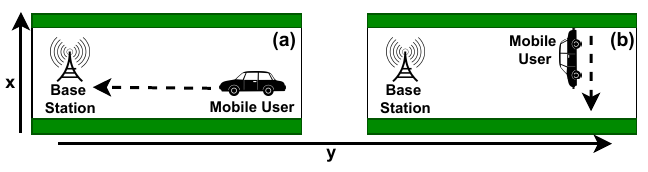}
    \centering
       \includegraphics[scale=0.30]{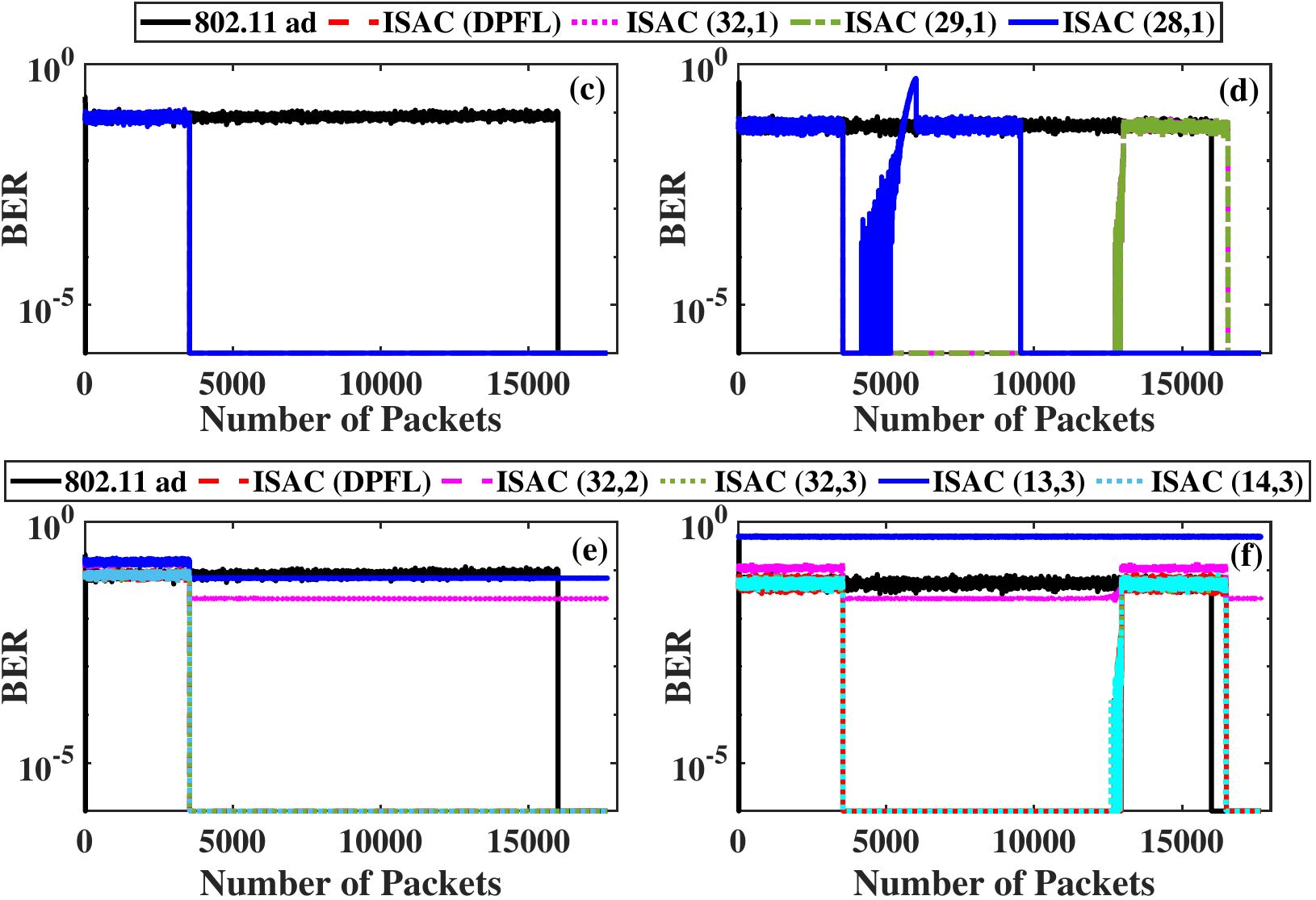}
    \caption{(a) Radial and (b) Transverse trajectory followed by \ac{MU} with respect to BS; \ac{ISAC} performance with fixed-point \ac{RSP} on \ac{FPGA} for a target in (c) transverse and (d) radial trajectories; \ac{ISAC} performance with a fixed-point transmitter on \ac{FPGA} for a target in (e) transverse and (f) radial trajectories. The comparison is with DPFL \ac{ISAC} and IEEE 802.11ad in Rician channel.}
    
    \label{fig:trajectory}
\end{figure}

In this section, we study how the radar performance metrics affect the subsequent communication metrics of the ISAC. We study the ISAC performance under a dynamic scenario where a point target MU travels along two distinct trajectories, radial and transverse, as shown in Fig.\ref{fig:trajectory}(a) and (b), respectively. In the radial trajectory, the MU moves in a radial direction away from the ISAC BS. Here, once the best beam at the BS is determined, the target remains within this beam during the remaining duration of the motion.  In the tangential trajectory, the target moves tangentially before the ISAC BS, as shown in Fig.\ref{fig:trajectory}(b). Due to the nature of the motion, the target frequently moves out of the main lobe of the BS, and the beam alignment has to recur often. Many strategies can be adopted for initiating a realignment, such as periodic or performance-based realignment. This work considers a setup where the realignment is initiated when the communication SNR falls below 15 dB. The performance metric is the resultant BER as a function of the number of packets (or time) during which beam realignment may happen multiple times due to the movement of \ac{MU} and we assume Rician channel conditions.

Figures~\ref{fig:trajectory}(c)-(d), compare the effect of WL of \ac{RSP} on the performance of \ac{ISAC} with DPFL WL-based transceiver. As shown in Fig.~\ref{fig:trajectory}(c), the initial \ac{BER} during the beam alignment phase is high when the target moves in radial trajectory. Thereafter, \ac{BER} is low, and further beam realignment is not triggered since \ac{MU} remains in the beam for the rest of the time. Interestingly, though the \ac{RMSE} of \ac{RSP} with WL of (28,1) is high, overall \ac{ISAC} performance is not degraded since the beam realignment is sufficiently accurate to get the low \ac{BER}. It can be observed that the proposed \ac{ISAC} offers faster beam alignment than the standard, resulting in a 5\% improvement in throughput. Figure~\ref{fig:trajectory}(d) considers a transverse trajectory where frequent beam realignment is necessary. It can be observed that the performance of \ac{RSP} with WL of (28,1) is poor, resulting in the frequent need for beam realignment compared to WL of (29,1). Even in this case, the proposed \ac{ISAC} offers better performance with  166.8\%  improvement in throughput. Similarly, Figs.~\ref{fig:trajectory}(e)-(f) compare the performance of \ac{ISAC} for different WL of the transmitter for radial and transverse trajectories, respectively. As expected, BERs for WL of (32,3) and (14,3) are nearly identical to that of DPFL. 

\subsubsection{Effect of Azimuth Precision}
\begin{figure}[!b]
\centering
\includegraphics[scale=0.4]{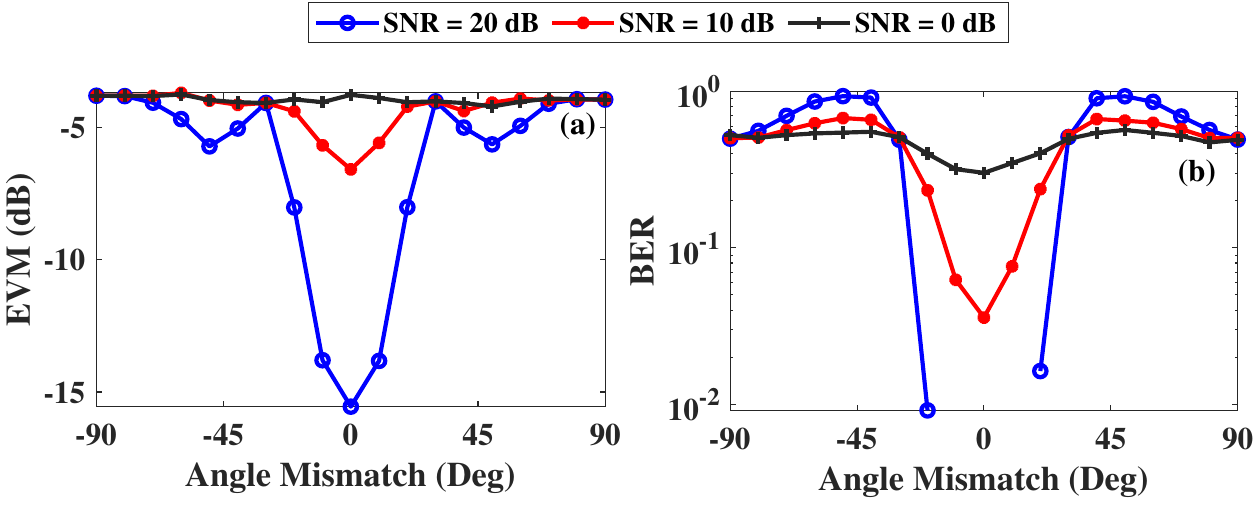}
\caption{Performance of \ac{ISAC} system (a) \ac{EVM}. (b) \ac{BER}}. 
\label{fig:Angle Mismatch}
\end{figure}

We study the variation of \ac{EVM} and \ac{BER} as a function of the beam misalignment between the BS and MU in Fig.~\ref{fig:Angle Mismatch}.  
The results indicate that as the angle mismatch between the BS and MU beams decreases, the \ac{EVM} and \ac{BER} improves, and the amount of improvement depends on the SNR. For an SNR of 20 dB, the maximum angle mismatch that can be accommodated is $10\degree$. However, even a slight misalignment increases both \ac{EVM} and \ac{BER}, and the EVM value degrades below the limit for 5G set by the 3GPP standard. The effect of beam misalignment is more pronounced at 10dB SNR, where a higher azimuth precision of $4\degree$ is required. However, at a low SNR of 0 dB, the system's performance is limited by noise. These results indicate the importance of beam alignment for \ac{mmWave} communications and how azimuth precision of radar eventually affects communication performance in ISAC. From RSP architecture perspective, though higher precision offers accurate beam alignment (i.e., lower azimuth \ac{RMSE}) and improved \ac{BER}, its execution time is higher.
For intance, the decrease in azimuth precision from $1\degree$ to $2\degree$ and $4\degree$ leads to improvement in \ac{RSP} execution time by 24\% and 56\%, respectively, allowing more time for data communication and, hence, higher throughput. In this paper, as we have 32 antenna elements at the BS, 4 antenna elements at MU, and the system is operating at mmWave frequency of 25 GHz, the beamwidth is 3.63\degree and 24\degree at BS and MU respectively.

\begin{figure}[!t]
    \centering
    \includegraphics[scale=0.35]{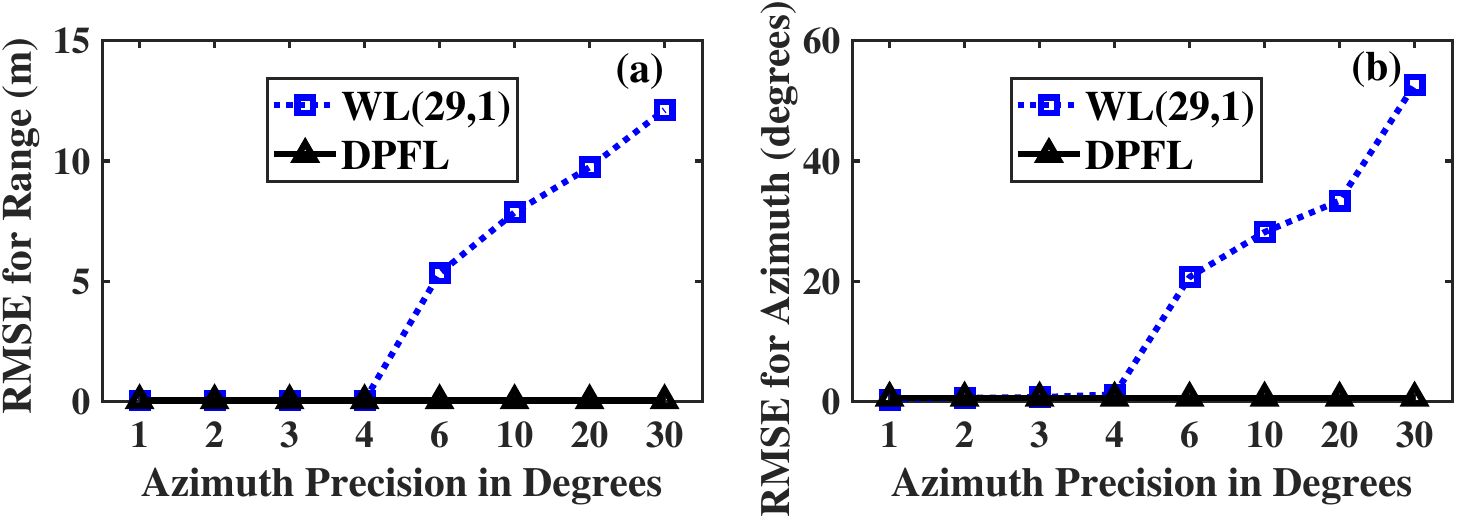}
    \caption{Performance analysis of \ac{RSP} for different azimuth precisions and WLs in terms of \ac{RMSE} for (a) range(m), and (b) azimuth (degrees).}
    \label{fig:PerformAnal}
\end{figure}

The proposed RSP architecture in Fig.~\ref{fig:RSP} can be reconfigured on the fly to change the azimuth precision from $1\degree$ to $10\degree$. However, we observed that the accuracy of fixed-point RSP architecture may not be sufficient to support the dynamic range required for lower precision.   As shown in Fig.~\ref{fig:PerformAnal}, the proposed RSP architecture with WL of (29,1) can support the azimuth precision of up to $4\degree$. To support a lower precision of $10\degree$, a further increase in WL is needed, leading to higher complexity. Thus, even though $10\degree$ precision is sufficient for an SNR of 20 dB, the RSP architecture with $4\degree$ must be used. For the radial and transverse trajectories considered in Fig.~\ref{fig:trajectory}, reconfigurable architecture that adapts azimuth precision based on SNR improves the throughput from 5\% and 166.8\% to 6.5\% and 171.2\%, respectively. 

\section{Proposed Architecture of \ac{MFE} with RF Impairments}
\label{sec:PA_MMFE}
\begin{figure}[!h]
\centering
\includegraphics[scale=0.3]{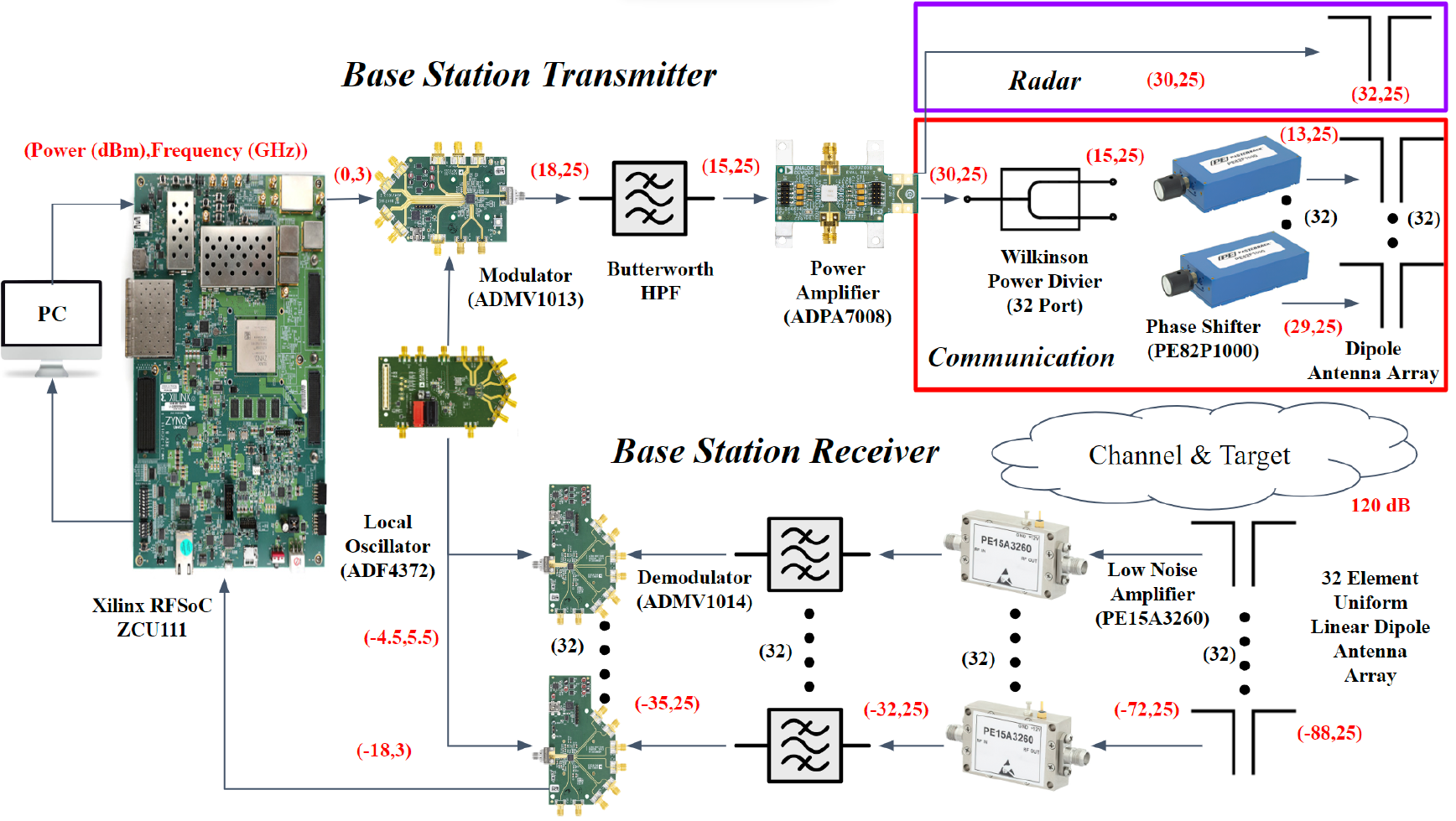}
\caption{Hardware level representation of MFE. Transmitter consists of a single dipole antenna for stage I (purple) and antenna array for stage II (red). The numbers within parentheses $()$ indicate (Power(dBm), Frequency(GHz)).}
\label{fig:Radar_AFE_act}
\end{figure}

This section presents the design details and architecture of a realistic \ac{MFE} implemented with \ac{RF} impairments on Mathworks Simulink R2022b.
The IEEE 802.11ad-based wideband \ac{ISAC} system hardware shown in Fig.~\ref{fig:Radar_AFE_act} is time multiplexed between radar-based sensing for detection and localization of \ac{MU} during stage I and communication with \ac{MU} during stage II. 
\subsubsection{Radar MFE} We first discuss the \ac{MFE} from the radar perspective. 
After digital modulation, the digital baseband radar signal of 2.64 GHz bandwidth is converted to analog through the digital-to-analog converter (DAC).  Subsequently, this signal is upconverted to an \ac{IF} of 3 GHz using an \ac{IF} modulator modeled after the corresponding Xilinx ZCU111 \ac{RFSoC} board component. The signal is then upconverted to 25 GHz using the \ac{IQ} modulator modeled after Analog Device's ADMV1013. Alongside modulation, this component provides a gain of 18 dB and has a 1dB gain compression point at 20 dBm. This is followed by a Butterworth high pass filter with a cut-off frequency of 10 GHz and a loss of 3 dB to the input signal. The filtered signal is provided to a high-power amplifier modeled after Analog Device's ADPA7008, with a gain of 15 dB and a 1 dB gain compression point at 31 dBm.
The amplified signal is provided to a half-wavelength dipole antenna centered at 25 GHz with a bandwidth of 3 GHz and a gain of 2.02 dBi. Though an antenna array is available at the \ac{BS} transmitter, only a single element is exploited for omnidirectional radar-based transmission since the angular position of the MU is not known at this stage. 
Next, we incorporate the two-way propagation loss ranging from 142 to 162 dB from the transmitter to a target/\ac{MU} modeled as an isotropic point scatterer of 0 dBsm radar cross-section and back to the receiver. 

The first stage of the radar receiver has a 32-element ULA, with the antenna elements spaced half a wavelength apart. Each antenna element is a half-wavelength dipole antenna identical to what was used during transmission. The \ac{ULA} introduces a gain of 16 dBi in the received signal. The self-interference from the transmitter is mitigated using the signal cancellation method as described in \cite{o1963high,lin2006kaband}. This is a commonly adopted strategy in radars with long duty cycle waveforms such as continuous wave or frequency modulated continuous waveforms. 


The received signal at each array element is amplified using an \ac{mmWave} low-noise amplifier modeled after Pasternack's PE15A3260, with a gain of 40 dB, a noise figure of 2.5 dB and a 1dB gain compression point at 13 dBm. The signal is then passed through an image rejection Butterworth high-pass filter with a cutoff frequency of 10 GHz and loss of 3 dB to remove harmonics outside of the band of interest. The \ac{mmWave} signal is downconverted from 25 GHz to the \ac{IF} of 3 GHz using an \ac{IQ} demodulator modeled after Analog Device's ADMV1014 with a gain of 17 dB, a noise figure of 5.5 dB, and a 1dB gain compression point of -14 dBm. The signal is then further downconverted to the baseband using the \ac{IF} demodulator based on the component on the Xilinx ZCU111 \ac{RFSoC} board. Finally, the signal is digitized with an analog-to-digital converter (ADC) with a sampling frequency of 5.28 GHz. As described in the previous sections, the \ac{RSP} is performed on the digitized received signal. Once the \ac{MU} is detected based on the target's \ac{SNR} and localized, the angular position of the \ac{MU} is used in the subsequent communication stage II. Based on these real-world component models used in the radar \ac{MFE}, the overall noise figure of the receiver system is 2.5 dB and a maximum detectable range of 41.22 m based on the Friis two-way propagation model. This matches the requirements of an ultra-short-range automotive radar for applications such as parking assistance, lane change assistance, and object recognition.
\subsubsection{Communication MFE} The \ac{ISAC} system's communication (stage II) starts immediately after radar functionality's (stage I) completion. 
Note that the same hardware components from Stage I are reused in Stage II. Here, the communication signal is \ac{OFDM} modulated in the DFE and converted to analog at the DAC, then upconverted to \ac{IF} and, subsequently, \ac{mmWave} in the \ac{MFE}. The bandpass filters remove intermodulation products and harmonics. 
The key difference between the radar and communication transmissions, is that a 32-port Wilkinson power divider feeds the \ac{mmWave} signal to the antenna array for communications, which introduces an additional loss of 15 dB. This antenna array is the same 32-element \ac{ULA} used in the receiver. However, on the transmitter side, this \ac{ULA} is configured to support analog beamforming by using \ac{mmWave} phase shifters at the transmission channel connected to each antenna array element. These phase shifters are modeled after Pasternack's PE82P1000 and have a loss of 2 dB. Then, we incorporate the path loss corresponding to one-way propagation between the \ac{BS} and \ac{MU} to the transmitted signal. 

The receiver on the \ac{MU} is a pure communications system (unlike the \ac{BS} with the ISAC system) with a four-element \ac{ULA} with elements spaced half a wavelength apart with a gain of 4 dBi. Each antenna is an omnidirectional dipole identical to the one used for \ac{BS}. The received signal again undergoes analog beamforming using phase shifters in the \ac{mmWave} chain connected to each receiver antenna, followed by a Wilkinson power combiner where all the signals are combined with a loss of 5 dB. Subsequent downconversion stages to \ac{IF} and baseband are identical to the radar receiver. This summarises the system-level implementation of the MFE. 
\subsubsection{\ac{RF} Impairments}
\label{sec:RF Impairments}
Under ideal conditions, none of the \ac{MFE} components will introduce distortions, and the radar and communication performances will be entirely governed by noise and channel conditions. However, in real-world situations, the \ac{RF} impairments in each component cause the overall performance to deteriorate.
We model the effects of different \ac{RF} impairments in the \ac{MFE} components through Simulink. The first set of impairments is the gain and phase imbalance between the modulator's I and Q channel ports at the transmitter. We model these imbalances by varying the gain disparity between the I and Q ports from 0 dB (ideal) to 9 dB (maximum imbalance) and the phase imbalance between the ports from the ideal 0\degree{} (ideal) to 30\degree{} (maximum imbalance). 

The \ac{IQ} modulator consists of a \ac{LO} that generates an \ac{mmWave} carrier frequency, which, along with the \ac{IF} signal, is fed to a mixer to generate a modulated \ac{mmWave} signal. In ideal conditions, the signals at the \ac{LO} and \ac{mmWave} ports are well isolated to prevent the occurrence of additional harmonics and intermodulation products. However, in real-world situations, some of the output from the \ac{LO} leaks into the \ac{mmWave} output port. A similar impairment can also be observed in the \ac{IQ} demodulator, where the \ac{LO} signal may leak into the \ac{mmWave} output port. We study the effect of these impairments by modeling the \ac{LO} after Analog Device's ADF4372 where the isolation between the \ac{mmWave} and \ac{LO} ports is limited to 12 dB. 

\begin{figure}[htbp]
\centering
\includegraphics[scale=0.5]{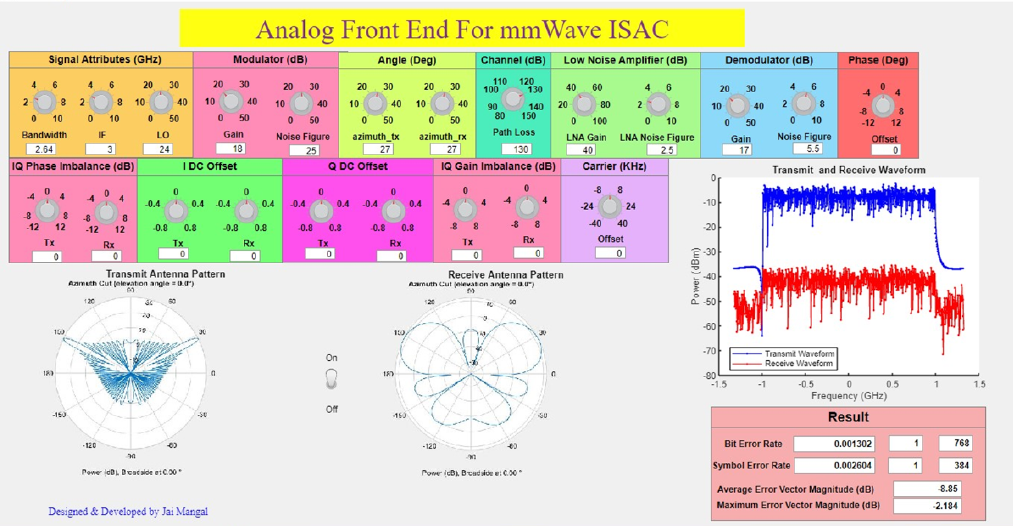}
\caption{Designed GUI of MFE ISAC system in Simulink R2022b.}
\label{fig:GUI}
\end{figure}

In real-world systems, the carrier frequency generated by the \ac{LO} at the \ac{IQ} demodulator in the receiver may be offset in terms of frequency and phase from the carrier frequency generated by the \ac{LO} at the \ac{IQ} modulator of the transmitter. Based on the specifications for ADF4372, we incorporate a maximum frequency offset of 100 MHz between the transmitter and receiver and compute communication performance results like \ac{BER} and \ac{EVM} and the estimates of the angular position of MU. The designed GUI of the \ac{MFE} in Simulink R2022b, through which system parameters can be controlled, is presented in Fig.~\ref{fig:GUI}. The GUI computes communication performance results like \ac{BER} and \ac{EVM} and the estimates of the angular position of MU.
\section{Results: Performance Analysis of Non-Ideal \ac{MFE}}
\label{sec:Results_MMFE}
We present both the communication and radar-based performance metrics for a realistic \ac{mmWave} front end. A key additional study is analyzing how radar performance subsequently affects the communication metrics in ISAC. 
\subsection{Communication System}
In the proposed 802.11ad \ac{ISAC}, the data are QPSK modulated at the transmitter. 
The received data symbols are mapped into a two-dimensional Euclidean constellation space with the I and Q components resolved along the X and Y axes, respectively. For an ideal \ac{MFE} without noise and impairments, the data symbols are mapped to $+45\degree, 135\degree, -135\degree,$ and $-45\degree$. However, besides path loss, the received data undergo corruption due to noise and hardware impairments, distorting the constellation phase diagram. Since the data bits are retrieved based on decision boundaries drawn around the four positions in the constellation, these distortions ultimately deteriorate the following performance metrics: (1) The \ac{EVM} or distance from the deviated position of the corrupted symbol on the phase diagram from the ideal position of an uncorrupted symbol; and (2) \ac{BER}.  
\begin{figure}[htbp]
\centering
\includegraphics[scale=0.4]{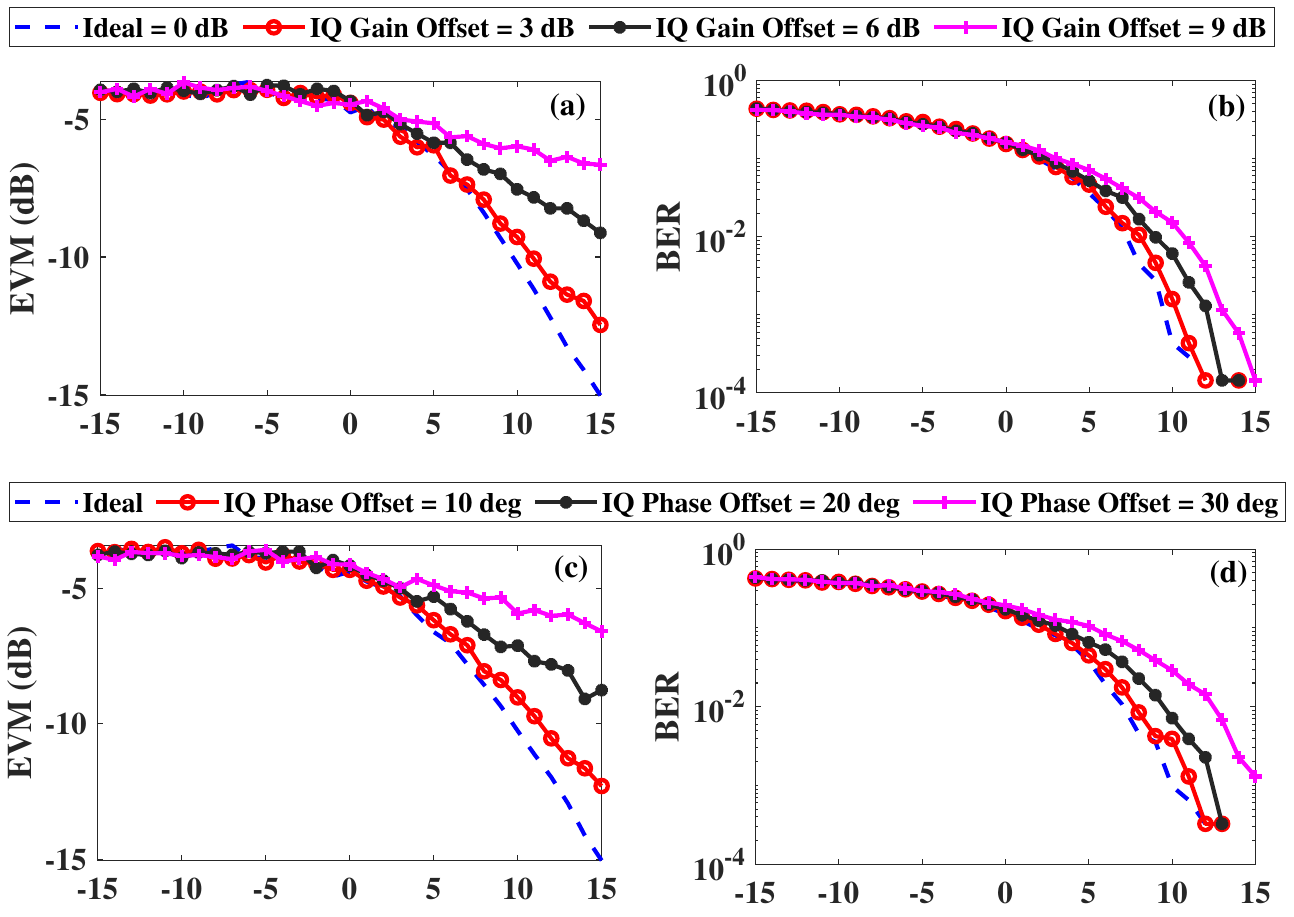}\\
\vspace{2mm}
\includegraphics[scale=0.4]{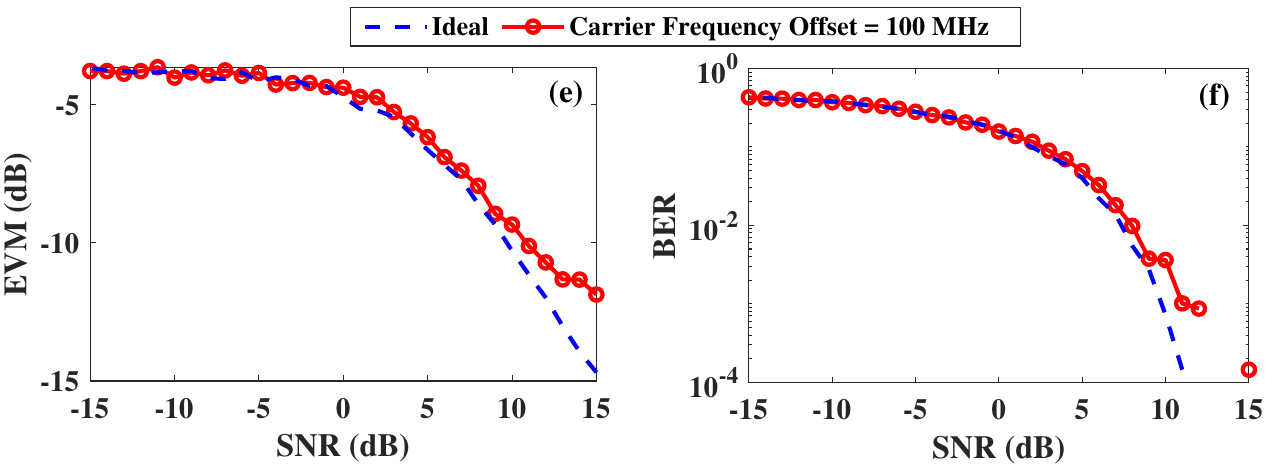}
\caption{Performance of communication system with the \ac{MFE} in terms of (a) \ac{EVM} v/s \ac{SNR}. (b) \ac{BER} v/s \ac{SNR}.} 
\label{fig:Communication Results}
\end{figure}

The \ac{EVM} and \ac{BER} of the communication system under \textbf{\emph{\ac{IQ} gain imbalance}} are illustrated in Fig.~\ref{fig:Communication Results}(a) and (b). The ideal \ac{MFE} exhibits an \ac{EVM} approaching -16 dB and a near zero \ac{BER} at an \ac{SNR} of 15 dB. With a fall in the SNR, the data symbols drift out of decision boundaries, increasing the \ac{EVM} and \ac{BER}. The problem is further exacerbated with an increase in \ac{IQ} gain imbalance, which causes the symbols to systematically shift closer to either the I or Q axis, depending on whichever is the stronger component. When the imbalance is very high, the symbols drift outside the decision boundary, resulting in an error in the data bit retrieval. 

Next, the \ac{EVM} and \ac{BER} of the communication system under \textbf{\emph{\ac{IQ} phase imbalance}} are depicted in Fig.~\ref{fig:Communication Results}(c) and (d). As the phase imbalance between the I and Q increases, the orthogonality between the components is not maintained, resulting in the diagonal drift of the symbols towards or away from their ideal positions in the phase diagram. As a result, the \ac{EVM} increases with the rise in phase imbalance. 
However, the \ac{BER} is low even up to an \ac{IQ} phase imbalance of 30\degree. This resilience can be attributed to the substantial decision boundary of $45\degree$ between symbols in the constellation phase diagram of a QPSK modulation scheme.  

Next, we study the impact of \textbf{\emph{\ac{CFO}}} between the transmitter and receiver LO on \ac{EVM} and \ac{BER} in Fig.~\ref{fig:Communication Results}(e) and (f). This offset induces spurs and phase noise into the signal, complicating accurate signal retrieval at the receiver. While an ideal \ac{MFE} will have zero offset, the realistic \ac{MFE} with ADF4372 LO can have a maximum offset of 100 MHz. The results show that \ac{EVM} and \ac{BER} are impacted significantly by the carrier offset mostly under high SNR conditions where the received constellation points continuously shift circularly in a clockwise or anticlockwise direction with respect to ideal constellation points. Under lower SNR, the results show that the performance is limited by noise rather than impairments. 

Finally, we examine the effect of \textbf{\emph{limited isolation between the LO and RF ports}} in the modulator or demodulator on \ac{EVM} and \ac{BER}. Here, we compare the metrics for an ideal \ac{MFE} with complete isolation between the two ports and a realistic \ac{MFE} with limited isolation of 12 dB corresponding to component ADF4372. Usually, when the isolation is poor, the constellation points shift circularly in a clockwise or anticlockwise direction with respect to the ideal constellation points. In our case, both the \ac{EVM} and \ac{BER} curves for the realistic \ac{MFE} mirror the ideal curves closely, indicating that 12 dB isolation is sufficient for practical purposes.

We benchmark the results of the proposed IEEE 802.11ad-based ISAC system with 5G standards defined by 3GPP. The EVM limit for 5G communication for QPSK modulation is 24.7\%. 
Figure~\ref{fig:Communication Results} shows that for an ideal MFE system, this EVM limit is reached at an SNR of 12 dB. However, when non-idealities such as an IQ gain imbalance of 3 dB, and IQ phase imbalance of $10\degree$, and \ac{CFO} of 100 MHz are incorporated in the MFE, the EVM limit is reached at an SNR of 15 dB. However, the LO to RF isolation of 12 dB provides a comparable performance with respect to the ideal MFE. These results highlight the importance of the proposed MFE for realistic link budget analysis. 

\subsection{Radar System}
The functional performance of the ISAC transceiver's radar component is illustrated in Table~\ref{tab:Angle_Range RMSE}. The evaluation focuses on the \ac{RMSE} for both the range and azimuth of the target, computed through \ac{RSP} across 400 Monte-Carlo simulations. The analysis considers four key \ac{RF} impairments: \ac{IQ} gain imbalance of 10 dB, \ac{IQ} phase imbalance of 10\degree, \ac{LO} to \ac{RF} output isolation of 12 dB, and \ac{CFO} of 100 MHz based on the specifications of the hardware components discussed in the earlier sections.
\begin{table}[htbp]
\centering
\caption{Angle and Range RMSE of Proposed ISAC}
\label{tab:Angle_Range RMSE}
\begin{tabular}{|c|c|c|c|c|}
\hline
\textbf{MFE} & \multicolumn{2}{c|}{\textbf{Angle RMSE}} & \multicolumn{2}{c|}{\textbf{Range RMSE}} \\ 
\hline 
\textbf{SNR} & \textbf{-5 dB} & \textbf{-10 dB} & \textbf{-5 dB} & \textbf{-10 dB} \\ 
\hline
Ideal & 0.45 & 18.9 & 0.08 & 1.62 \\
\hline
IQ Gain Offset & 0.47 & 23.25 & 0.08 & 0.52 \\
\hline
IQ Phase Offset & 0.46 & 19.35 & 0.08 & 0.63 \\
\hline
LO to RF Leakage & 0.43 & 30.55 & 0.08 & 1.02 \\
\hline
Carrier Frequency Offset & 0.31 & 32.73 & 0.08 & 0.77 \\
\hline
\end{tabular}
\end{table}
The examination of the results reveals that, even up to a low \ac{SNR} of -5 dB, the radar robustly estimates the target range and azimuth despite the impairments. However, the radar becomes more sensitive to the impairments at a low SNR of -10 dB. 
\subsection{Impact of radar performance on communications}
\begin{figure}[htbp]
\centering
\includegraphics[scale=0.4]{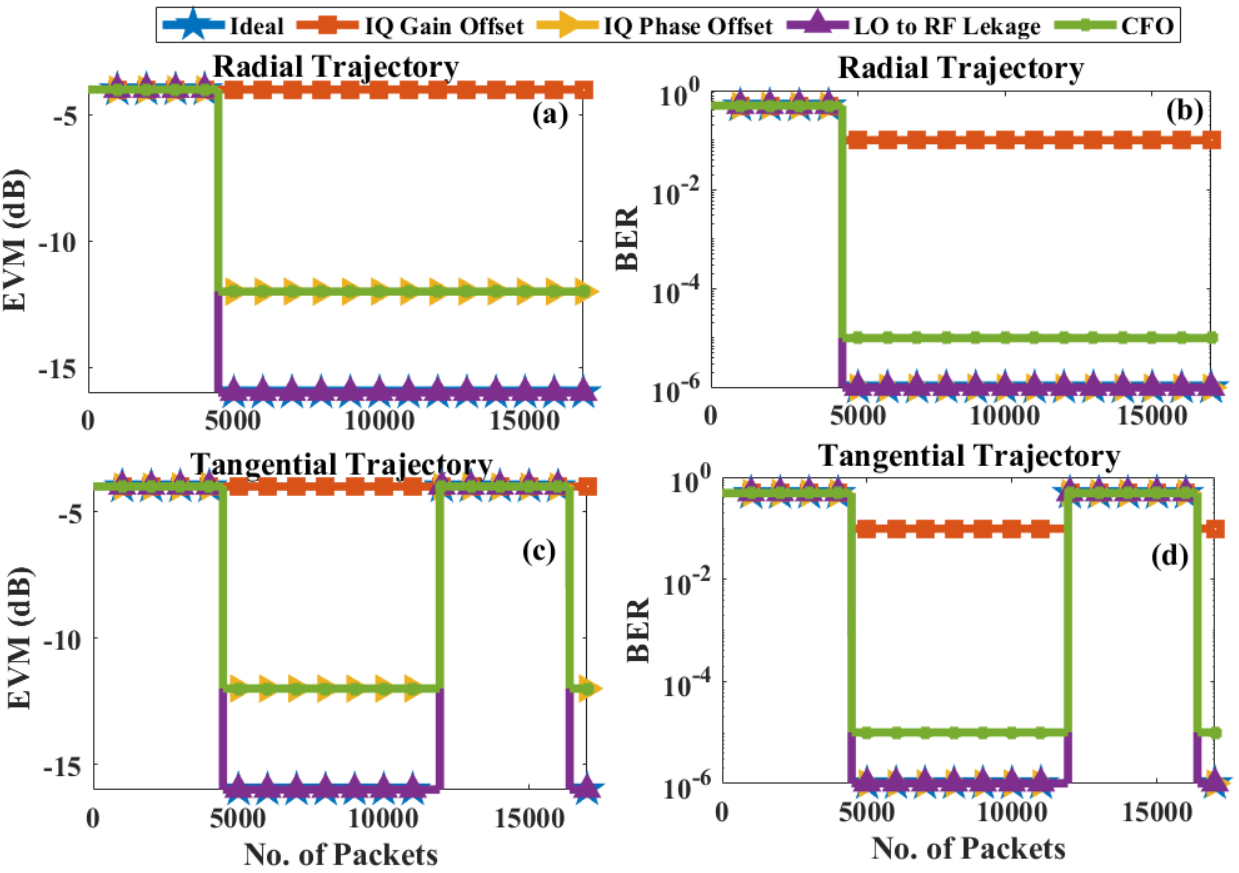}
\caption{Performace of \ac{ISAC} system on different trajectories (a) \ac{EVM} for radial. (b) \ac{BER} for radial. (c) \ac{EVM} for tangential. (d) \ac{BER} for tangential.}
\label{fig:JRC Trajectory Results}
\end{figure}

In this section, we repeat the experiments in Section~\ref{Sec:ISACFPGA} for different RF impairments. Along with BER, we also consider EVM as performance metric. As shown in Fig.~\ref{fig:JRC Trajectory Results}(a) and (b), once the MU's angle is determined, the ISAC BS transmitter switches to high gain directional transmission, resulting in a subsequent fall in both \ac{EVM} and \ac{BER} with high SNR. In this trajectory, radar functionality is not resumed since the MU and BS remain aligned during the remaining duration. It can be observed that the fall in the error metrics after beam alignment is very significant for an ideal MFE. However, that is not the case when there are RF impairments. Specifically, the results show that the overall system performance is most sensitive to the effect of the IQ gain imbalance and least sensitive to the \ac{CFO}.

In the tangential trajectory case, 
Figures~\ref{fig:JRC Trajectory Results}(c) and (d) depict the performance metrics of \ac{EVM} and \ac{BER} over the course of time/number of packets. The graph shows that these error metrics are initially high due to the initial omnidirectional transmission by the BS. After the completion of \ac{RSP}, the MU's angle is determined, and directional communication with high-gain narrow beams commences, resulting in a fall in the error metrics. The errors again become poor during the MU's motion due to the misalignment of the beams. Again, the results show the sensitivity of the overall system performance to the effects of RF impairments, especially due to the IQ gain imbalance. 


The throughput of the proposed ISAC with an ideal \ac{MFE} is 0.77 Gbps and 0.66 Gbps for radial and tangential trajectory, respectively. In the presence of RF impairments due to non-ideal MFE, we observed a degradation of 8.57\% and 6.61\% for radial and tangential trajectory, respectively. Despite this degradation, the 802.11ad-based ISAC offers 34.84\% and 50.12\% higher throughput for radial and tangential trajectories, respectively, than the IEEE 802.11ad standard with ideal MFE.
\section{Conclusion}
\label{sec:Conclusion}
We designed and implemented a baseband DFE of an IEEE 802.11ad-based \ac{ISAC} system on FPGA integrated with a wideband MFE in MATLAB's Simulink. The DFE's performance is validated for different wireless channels, target trajectories, and word lengths and significantly improves throughput over the conventional IEEE 802.11ad standard. We also demonstrated the reconfigurable architecture for the on-the-fly selection of azimuth precision. Based on the free space link budget analysis, we estimate a maximum detectable range of approximately 41m for the system, which meets the requirements of ultra-short-range radars commonly used in many automotive radar applications. We also study the interrelated radar and communication metrics. Any deterioration of the beam alignment above $10\degree$ due to poor radar performance will result in an overall degradation in the communication EVM below the 3GPP recommendations. Further, RF impairments such as IQ gain and phase imbalance, \ac{CFO}, and RF isolation deteriorate the overall system throughput between 6 to 8\%. Despite this degradation, the 802.11ad-based ISAC offers significant improvement (above 34\%) in the overall throughput compared to the IEEE 802.11ad standard with ideal MFE. 
Future work will involve developing and testing the MFE hardware prototype for ISAC on real radio channels.


\bibliographystyle{ieeetr}
\bibliography{reference}

\end{document}